\documentclass[final]{siamltex}

\usepackage{url,amsmath,amsfonts,amssymb,paulmath, fullpage}

\newcommand{\di}{\displaystyle}
\newcommand{\al}{\alpha}

\newcommand{\be}{\beta}

\newcommand{\D}{\mathcal{D}}
\newcommand{\cE}{\mathcal{E}}
\newcommand{\un}{\underline}

\title{Codes from Zero-divisors and Units in Group Rings}

\author{Paul Hurley\thanks{IBM Research, Zurich Research Laboratory,
Switzerland. pah@zurich.ibm.com}
\and
Ted Hurley\thanks{National University of Ireland, Galway, Ireland.
ted.hurley@nuigalway.ie}
}

\begin{document}
\maketitle

\begin{abstract}
We describe and present a new construction method for codes using encodings from
group rings. They consist primarily of two types: zero-divisor and
unit-derived codes. Previous codes from group rings focused on ideals;
for example 
cyclic codes are ideals in the group ring over a cyclic group. The fresh
focus is on the encodings themselves, which only under very limited
conditions result in ideals.

We use the  result that a group ring is isomorphic to a certain
well-defined ring of matrices, and thus every group ring element has an
associated matrix. This allows matrix algebra to be used as needed in
the study and production of codes, enabling the creation of standard generator
and check matrices.

Group rings are a fruitful source of units and zero-divisors from which
new codes result. Many code properties, such as being LDPC or self-dual,
may be expressed as properties within  the group ring thus enabling
the construction of codes with these properties.
The methods are general enabling the construction of codes with 
many types of group rings. There is no restriction on the ring and
thus codes over the integers, over matrix rings or even over group
rings themselves are possible and fruitful.
\end{abstract}

\section{Introduction}
We present techniques for construction of codes from encodings in group rings,
resulting in two new types of codes that we call {\em zero-divisor codes} and
{\em unit-derived codes}. Many existing codes use zero-divisors of
special group
rings. All cyclic codes, such as BCH, Golay, Hamming and Reed-Solomon are
special types of zero-divisor codes of the group ring of the cyclic group.

The use of group rings for the construction of codes has, to date, been
concerned with the ideals contained within. As cyclic codes are ideals in the
group ring over the cyclic group, this has lead to considering the natural
generalisation of cyclic codes as being ideals
(e.g.~\cite{macwilliams69-codesfromideals}).  Indeed, a group ring code has been
defined (e.g.~\cite{hughes_2000}) as an ideal in a group ring. 
A group ring is often called  a group algebra when the ring is a field.

Codes from group ring encodings, presented here, are submodules in the
group ring and only in certain restrictive cases are they ideals
in a group ring. Indeed the unit-derived codes are never ideals. The
methods for obtaining generator and check matrices apply in all cases.

We use an isomorphism between a group ring and a certain ring of matrices
to enable the straightforward production of check and generator matrices for the
codes directly from zero-divisors and units. The interplay between the group
ring construction and the corresponding matrix construction is extremely
fruitful, allowing immediate calculation
(algebraically) of generator and check matrices for these codes.

Looking at group ring encodings expands the space of possible codes, and, 
we believe, offers a simple and intuitive approach

Properties of codes from group ring encodings, such as self-duality or
having a sparse parity-check matrix (namely LDPC) often have an 
easy algebraic description as properties in zero-divisor or
unit-derived  codes within
group rings and this description can be exploited for the construction and
analysis of these examples. We present a number of examples of these LDPC
 and self-dual  zero-divisor and unit-derived codes. 

It is also possible to obtain convolutional codes from unit-derived and
zero-divisor codes from group rings
and this is the subject of a companion work.
Also of note is that the underlying algebraic structure of group rings
often allows the calculation of distance directly.

The explicit examples presented are necessarily of a short length. The
methods, however, are completely general and can be used to construct codes with longer lengths. The algebra can then be used to derive
properties of these codes or to construct long length codes with some desired property.
 
\subsection{Overview and layout}
We begin, in \sref{grs_as_rings}, with a description of the  basic
algebra, defining a group ring $RG$ and zero-divisors and units.  
We use the result that  $RG$ is isomorphic to a certain ring of
$n\times n$ matrices over the ring $R$ where $n$ is the order of the group
$G$ -- see
\cite{hurley_grouprings2006} . These matrices are called {\em
  $RG$-matrices} over $R$. 
This connection is used later to derive generator and check matrices.
Examples of  $RG$-matrices for different groups $G$ are given -- these include
circulant matrices, block-type circulant matrices, Toeplitz-type
matrices, and Walsh-Toeplitz matrices.

\sref{encodingdefns} defines a code from a group ring encoding. In
\sref{zerodivisors}, the procedure for obtaining zero-divisor codes is
presented, showing how generator and check matrices may be obtained.

Then the concept of unit-derived codes, which are
particularly elegant and simple, is described in
\sref{codesfromunits}. Obtaining generator and check matrices from
units is straightforward, there is much flexibility
from which to generate the code, and the invertibility of units offers
particular promise for exploitation.

The algebraic group ring descriptions enable new codes and series of codes
with particular properties or of a particular type to be constructed. The group ring description may often be used to  analyse, or deduce properties of, the codes. 

There is no restriction on the ring $R$ in the group ring $RG$ and
what is required is either a zero-divisor or a unit within the group ring.
Thus, the construction of codes over systems other than fields is possible. In
particular, integral codes together with their generator and check matrices may
be constructed using units in group rings over the integers.

Example codes are shown in \sref{examplecodes}, starting with some illustrative
self-dual codes.
Examples of codes when working in group rings over dihedral groups are then
given in \sref{dihedral} (for example an explicit construction of a
$(62,30,12)$ code, which is one of best known distance). Using
encodings directly and the $RG$-matrix structure, as shown in this
work, is new. Previous examinations of codes using dihedral groups worked
exclusively using ideals.

In \sref{ldpc2} we  describe examples of codes designed to have a sparse check
matrix -- LDPC (Low Density Parity Check) codes. 
Most techniques to date obtain LDPC codes
by randomisation, but there are recent developments using algebraic
techniques (which have advantages such as immediate generator matrices and
structure to exploit). We show that some of these implicitly operate in a group
ring. Then 
we propose a regular LDPC code from a group ring of  a direct  product of
groups - from either a zero-divisor or unit-derived code.

Relationships to well-known codes is explored in
\sref{relationships}. For example, we show that when we specialise to the
zero-divisor codes of group rings on cyclic groups which are ideals 
we get the {\em cyclic} or {\em polynomial} codes; one is not restricted to ideals
and so even in the group ring of the cyclic group, the
zero-divisors codes can define codes other than those which are  
termed {\em cyclic}.
The (restrictive) conditions under which one of these codes (for
general group rings) is an ideal in the group ring are also derived in this
section.

\section{Group rings}
\label{sec:grs_as_rings}
Only basic, necessary definitions and results on group rings are provided. For
further information on group rings and related algebra, please
see~\cite{milies_grouprings}.

Let $G$ be a group and $R$ a ring\footnote{The ring can be
arbitrary; in all cases dealt with here, the ring
$R$ has an identity.}.
The {\em group ring}
$RG$ is a ring consisting of the set of all summations $u =\di\sum_{g\in
G} \al_g g$, where $\al_g\in R$. Consider also $v=\di\sum_{g\in G}
\beta_g g$.  Addition is defined term-by-term,
$$u + v = \sum_{g\in G} (\al_g+\beta_g) g,$$ while multiplication is a
convolution-like operation,
$$ 
u v = \sum_{g,h\in G} \al_g\beta_h \,§ gh.
$$
This can be written, since $G$ is a group, as:
$$
u v = \sum_{g\in G} \biggl(\sum_{h \in G} \al_h\beta_{h^{-1}g}\biggr)g.
$$

For example, if $G$ is a finite cyclic group, multiplication is the
circulant convolution.

$G$ acts as a basis for the module $RG$ over
the ring $R$. 
In effect, every element of $RG$ is a vector composed
of elements of the ring $R$, with the $i$th component associated
with the group element $g_i$; addition is of components and
multiplication is obtained from the group multiplication together
with the distributive law.

The following definition applies in general to any ring
  but we shall be particularly interested in cases
  where the ring is a group ring.

\begin{definition}
For any ring $R$ a  non-zero element $z\in R$ is a {\em
zero-divisor}\footnote{Strictly speaking we have defined
 a left zero-divisor but in all situations here when an element is
a right zero-divisor it will also be a left zero-divisor.}  if and only if there
exists a non-zero $r\in R$ such that $zr = 0$.
For any ring $R$ with identity an element  $u\in R$ is a {\em unit} if
  and  only if there exists an
element $v \in R$ such that $uv=1 = vu$.
\footnote{In many  situations  $uv=1$ automatically implies $vu =1$.} 
When this $v$
exists it is often written as $u^{-1}$. 
\end{definition}

Sometimes  $R$ is a field in the group ring $RG$, but is not
restricted to  such, and then $RG$
is called a {\em group algebra}. However, other rings such as when $R=\Z$, the 
integers,  cases where $R$  has zero-divisors or where $R$ itself
is  a group ring, have also proved useful.

\subsection{Group rings and matrices}
\slabel{gr-and-matrices}
Let $\{g_1, g_2, \ldots , g_n\}$ be a fixed listing of the elements of
$G$. The {\em $RG$-matrix} of $w = \di\sum_{i = 1}^n\al_{g_i}g_i \in RG$, is
in $R_{n\times n}$, the ring of $n\times n$ matrices over $R$, and
defined as,
\begin{align}
\elabel{rgmatrix}
M(RG,w)=
\begin{pmatrix}
\alpha_{g_1^{-1}g_1} & \alpha_{g_1^{-1}g_2} & \ldots &
 \alpha_{g_1^{-1}g_n} \\ \alpha_{g_2^{-1}g_1} & \alpha_{g_2^{-1}g_2} &
 \ldots & \alpha_{g_2^{-1}g_n} \\ \vdots & \vdots &\vdots &\vdots \\
 \alpha_{g_n^{-1}g_1} & \alpha_{g_n^{-1}g_2} & \ldots &
 \alpha_{g_n^{-1}g_n}
\end{pmatrix}
\end{align}
The first column is, in essence, labelled by $g_1$, the second by
$g_2$ etc.; the first row is in essence labelled by $g_1^{-1}$, the
second by $g_2^{-1}$ etc. Each row and each column is a permutation, 
determined by the group multiplication, of the initial row. 

The first important result, proven in \cite{hurley_grouprings2006}, is that a
group ring $RG$ is isomorphic to a ring of {\em $RG$-matrices} over $R$, a
subring of $R_{n\times n}$, the ring of $n\times n$ matrices over $R$.
\begin{theorem}
\label{thm:main}
Given a listing of the elements of a group $G$ of order $n$ there is a
bijective ring homomorphism $\sigma: w \mapsto M(RG,w)$ between $RG$
and the $n\times n$ $RG$-matrices over $R$.\endproof
\end{theorem}

This result means that the group ring and the ring of matrices are
interchangeable. One can thus exploit results from matrix algebra and group
rings as needed.

For every $u\in RG$, its $RG$-matrix $\sigma(u)$ is denoted by the
corresponding capital letter $U$.

The next useful result, shown also  in \cite{hurley_grouprings2006}, is
that in a group algebra, every element must be a unit or zero divisor, and that
there is a method to determine which.
\begin{theorem}
\label{thm:groupalgebra}
Let $R$ be a field. A non-zero $u\in RG$ is a zero divisor if and only
if $\det(\sigma(u)) = 0$, and otherwise a unit. \endproof
\end{theorem}

Thus, when $R$ is a field, 
an element $u \in RG$ is a zero-divisor if and only if $\rank U <n$ and is a
unit if and only if $\rank U = n$.
Additionally, it can easily be shown that a finite ring with identity contains only zero-divisors and units.

The isomorphism between group rings and the $RG$-matrices 
allows the generator and check matrices for the group ring codes, to
be defined in \sref{gr_codes}, to be immediately derived from the group ring
description.

Many of the results stated here 
hold for infinite groups with corresponding infinite $RG$-matrices.

\subsubsection{Examples of $RG$-matrices}
 
In the cyclic group ring case the matrices are the {\em circulant}
matrices~\cite{davis79-circulantmatrices}.  All cyclic codes can be generated
from singular circulant matrices.

The $RG$-matrix types which turn up as isomorphic to
certain group rings include Toeplitz-type matrices, Walsh-Toeplitz
matrices and circulant matrices, Toeplitz combined with Hankel-type matrices and block-type circulant matrices.

In the general finite abelian group case, $RG$ is isomorphic to
certain block circulant matrices that, when $R$ is commutative,
commute and  are normal.
In the case of an elementary abelian 2-group of rank
$m$ and order $2^m$, where the matrix size is $2^m\times 2^m$, the
$RG$-matrices are the Walsh-Toeblitz matrices over $R$. In the
case of the dihedral group the $RG$-matrices are of the form
$\begin{pmatrix} A & B \\ B & A \end{pmatrix}$ where $A$
is a (general) circulant matrix and $B$ is 
a reverse circulant matrix.

Further details on these examples of $RG-$matrices appear
in~\cite{hurley_grouprings2006}.  

\subsection{Element properties}
Many concepts and properties of matrices turn out to have useful equivalents in the group ring context. These are inherent to the group ring itself, existing independent of any group listing chosen to establish the isomorphism, while maintaining consistency with their equivalent matrix evocation.

We first define the transpose of a group ring element.
\begin{definition}
The {\em transpose} of an element $u = \sum_{g\in G} \alpha_g g$ in $RG$ is $u\T = \sum_{g\in G} \alpha_g g^{-1}$, or equivalently, $u\T= \sum_{g\in G} \alpha_{g^{-1}} g$.
\dlabel{transpose}
\end{definition}

This is consistent with the matrix definition of transpose. For a given
listing $G=\{g_1,\ldots,g_n\}$, let $U$ be the $RG$-matrix of $u$. The entry
$(i,j)$ of the $RG$-matrix of $u\T$ is $\alpha_{(g_i^{-1} g_j)^{-1}} =
\alpha_{g_j^{-1} g_i}$, so $U\T$ is the $RG$-matrix of $u\T$.

The transpose $u\T$ has also been called the canonical
antiautomorphism of $u$~\cite{ward-qrcdivisibility}, denoted $\overline{u}$.
When dealing with the cyclic groups, it has been referred to as a
transpose~\cite{macwilliams71orthogonalcirculant}, and it is associated with
the reciprocal polynomial.

\begin{definition}
We say that $u\in RG$ is {\em symmetric} if and only if $u\T=u$.
\end{definition}

Clearly, the definition is consistent as $u$ is symmetric if and only if $U$
is a symmetric matrix.

\subsection{Some Notation}
We shall denote $Wu = \{xu: x\in W\}$ and $uW=\{ux:x\in W\}$.

The notation $\un{x}$ is used to indicate that $\un{x}$ is a vector as
opposed to an element of the group ring. For $\un{x}=(\alpha_1,\alpha_2
,\dots,\alpha_n)\in R^n$, the mapping $\zeta(\un{x}) = \sum_{i=1}^n
\alpha_i g_i = x$ is an element in $RG$ according to a given listing
of $G$.  $\zeta\I(x)$ denotes the inverse map.

\section{Codes from group ring encodings}
\slabel{encodingdefns}
\label{sec:gr_codes} Let $RG$ be the group ring of the group $G$ over
the ring $R$. A listing of the elements of $G$ is given by $G =
\{g_1,g_2,\ldots,g_n\}$. Suppose $W$ is a submodule of $RG$, and $u\in RG$ is
given.

\begin{definition} Let $x\in W$. A {\em group ring encoding} is a
mapping $f:W \rightarrow RG$, such that $f(x)=xu$ or $f(x)=ux$.  In
the latter case, $f$ is a {\em left group ring encoding}. In the
former, it is a {\em right group ring encoding}.
\end{definition}

A code $\C$ {\em derived from a group ring encoding} is then the image of a 
group ring encoding, i.e. for a given $u\in RG$, $\C=\{ux: x \in W\}$ or 
$\C=\{xu : x \in W\}$.

Multiplication need not necessarily commute in a group ring. Allowing
non-commutative groups enables the construction of non-commutative
codes.
\begin{definition}
If $xu = ux$ for all $x$ then the code $\{xu: x\in W\}$ is said to be {\em
commutative}; and otherwise {\em non-commutative}.
\end{definition}

When $u$ is a zero-divisor, it generates a {\em zero-divisor code} and
when it is a unit, it generates a {\em unit-derived code}.

There is no restriction on the ring $R$. It can be a field, but 
the techniques we describe are more general, and enable codes over other rings
such as the integers $\Z$, rings of matrices or others.

In practice, the submodule $W$ has dimension $r < n$. It can have the
basis $\{g_1,g_2,\ldots, g_r\}$.
Other submodules also turn out to be useful, e.g.
as generated by $\{g_{k_1}, g_{k_2}, \ldots, g_{k_t}\}$ with $1 \leq t < n$
where $\{k_1, k_2, \ldots, k_t\}$ is a subset of $\{1, 2, \ldots, n\}$.

For unit-derived codes, there is complete freedom in the choice of $W$ (and
hence $r$). Zero-divisor codes, as we show in \sref{zerodivisors}, have
restrictions placed on what $W$ can be in order for a one-to-one map from the code back to $W$ to exist.

When $RG$ is finite and has an identity, only zero-divisors and units are
contained in $RG$. This is also true when $R$ is a field by
\thmref{groupalgebra}. We note, in passing, that in other cases it is possible
to generate codes from group ring encodings which are neither zero-divisors
nor units, producing a so-called {\em non-zero divisor} code. Investigation of
the properties of such codes is under investigation.

\section{Codes from zero-divisors}
\slabel{zerodivisors} We now concentrate on constructing codes from
zero-divisors in a given group ring $RG$. 

Assume  $G$ is of order $n$
with listing $\{g_1, g_2, \ldots, g_n\}$. The code will be of length
$n$ and its dimension will depend on the choice of the submodule $W$.  The
presentation will first deal with details from the group ring, and
then incorporate the matrix algebra relationship.

Let $u$ be a zero-divisor in $RG$, i.e. $uv=0$ for some non-zero $v
\in RG$. Let $W$ be a submodule of $RG$ with basis of group elements
$S \subseteq G$.

As defined in \sref{encodingdefns}, a resultant zero-divisor code is
$\C=\{ux: x \in W\} = uW$ or $\C=\{xu : x \in W\}= Wu$. The code is thus
constructed from a zero-divisor $u$, a submodule $W$ and, for $RG$
non-commutative, a choice over left or right encoding. We shall
describe the case
of right-encoding, that is $\C = Wu$. The left-encoding case is
similar but here we need to consider $vu=0$.
 
We say that $u$ is a {\em generator element} of the code $\C = Wu$ relative
to the submodule $W$. It is of course possible that $\C$ has another
generator element and indeed may also be defined in terms of a
different submodule $W$. 

The case when $Wu = RGu$  is the particular traditional case where the
code is a left ideal -- see \sref{relationships} for a more complete
discussion on this. This is the case where $\rank U$ has the same rank
or dimension as $Wu$.

When $u$ is a zero-divisor then there is an element $v \not = 0$ with
$uv = 0$ and thus $y \in \C$ satisfies $yv = 0$. It may happen that 
 such an element $v$ which will also determine the code. 
\begin{definition}
$v\in RG$ is said to be a (left) {\em check element} for a
  zero-divisor code $\C$
when $y\in \C$ if and only if $vy=0$. We can then write $\C=\{y \in
RG: vy=0\}$.
\end{definition}

We shall show that given a zero-divisor $u$ and the code $\C$ 
there is a set $v_1, v_2,
\ldots , v_t$ of elements in $RG$ such that $y \in \C$ if and only if 
$yv_i = 0$ for $1\leq i \leq t$.

Zero-divisor codes with a single check element are particularly
useful and exist in many cases. 

 We have defined codes in $RG$ as generated by a zero-divisor and
 relative to a submodule $W$. 
We note that in addition to using a zero-divisor as a generator, codes
can also be constructed by using a zero-divisor instead directly as a 
check element,
regardless of whether it has a single generating element or not. 
\begin{definition}
Suppose $T$ is a submodule of $RG$. Define $T_v = \{x \in T | xv =
0\}$ and say $T_v$ is the {\em check zero-divisor code} relative to $T$.
\end{definition}

Note that $T_v$ is a submodule of $RG$ and in the case where $T= RG$
we have that $T_v$ is actually a left ideal. It only makes sense to
consider the case where $v$ is a zero-divisor in which case $T_v \not
0$. In some cases this code will have a single generator matrix but in
any case it will be possible to describe a set of generator elements, 

\subsection{Modules}
\slabel{choosing-basis}

Here we are now restricting our attention to the case when $R$ is a field. Some
of the results hold over integral domains and also for rings in
general but we do not deal with these complications here.

(In the unit-derived codes,
\sref{codesfromunits}, we do not have the same restrictions.).
\begin{definition}
A set of group ring elements $T \subset RG$ is {\em linearly
independent} if, for $\alpha_x \in R$, $\sum_{x \in T} \alpha_x x
=0$ only when $\alpha_x=0$ for all $x\in T$. Otherwise, $T$ is {\em
linearly dependent}.
\end{definition}
We define $\rank(T)$ to be  the maximum number of linearly
independent elements of $T$.  Thus $\rank(T)=|T|$ if and only if $T$ is
linearly independent.

Note that a zero-divisor code $\C= Wu$, where $W$ is generated by $S$, 
is the submodule of $RG$ consisting of all
elements of the form $\sum_{g \in S} \alpha_g gu$. The dimension of
this submodule is thus $\rank(Su)$. 

If $Su$ is linearly dependent then
there exist a subset $S'u$ of $Su$ which is linearly independent and
generates the same module as $Su$. \footnote{It is here  we require
 that $R$  be a field.} Let $W'$ to be the submodue of $W$
generated by  $ S'$  and then the code $\C = Wu =
W'u$, and $S'u$ is linearly independent.   

The maximum dimension a code for a given zero-divisor $u$ is 
$r=\rank(Gu)$. 

The zero-divisor codes are thus $(n,k)$ codes for where $k =
\rank(Su)$ and $k \leq r = \rank(Gu)$. As pointed out for a given $u$
and a given $W$ it is always possible to find a submodule $W'$ of $W$
which may be $W$ itself such that $W'$ is generated by $S'$ with $S'u$
linearly independent and $Wu = W'u$. 
One way of finding $S'$ inside $S$ is explained below using the matrix $U$
of $u$  and finding an appropriate basis for the matrix consisting of
the relevant rows of $U$ corresponding  to the elements of $Su$. 

A way of finding an $(n,t)$ zero-divisor code is to 
find  $t$ linearly independent rows 
$i_1, i_2, \ldots , i_t$ of $U$. Then $S= g_{i_1}, g_{i_r2},
\ldots, g_{i_t}$ is such that $Su$ is linearly independent and
generates an $(n,t)$ code. The case  $t = \rank
(U)$ is the code $RGu$ and can be obtained from considering (any) $t =\rank(U)$
linearly independent rows of $U$. The codes with $t < \rank(Su)$ can
be considered as `shortened' codes. Their generator and check matrices
are easly obtained by methods described below.

Notice that to say $Su$ linearly independent is equivalent to saying that
$W$ contains no zero-divisors of $u$.

For $G$ the cyclic group $C_n=\{1,g,g^2,\ldots,
g^{n-1}\}$, let $r$ be the first value
such that $\{u,gu,g^2u,\ldots,g^ru\}$ is linearly dependent. Then $r$
is the $\rank(u)$, and any subset of $
S= \{1,g,g^2,\ldots,g^{r-1}\}$ can be chosen to generate $W$. 
The proof of this is
straighforward and is given in
\sref{rank-proofs} below. Shortened cyclic codes are obtained by choosing a
subset of this set $S$. 

\subsection{Equivalent Codes in $R^n$}
The established relationship between the group ring and matrices
in~\sref{gr-and-matrices} enables us to express the code in terms of
matrices, and to derive generator and check matrices for an equivalent
code.  More precisely, we shall define two equivalent codes in the
module $R^n$, the second of which has generator/check matrices of the
usual form.

Let $U$ be the $RG$-matrix of $u$, and $W$ be a submodule with basis\\
$S = \{g_{i_1},g_{i_2}, \ldots ,g_{i_r}\}$ such that $Su$ is linearly
independent. 
As previously stated, $\C= Wu$ is a code defined on $RG$. A $(n,r)$ code
$\cE$ can be defined from $R^r$ to $R^n$ as follows. Let $\un{w} =
(\al_1,\al_2,\ldots,\al_r)\in R^r$ be the vector to be encoded. 
Using the basis $S$, we write $\un{w}$ as $\un{x}\in R^n$ with $\al_j$
in position $i_j$ for $1\leq i \leq r$ and zero everywhere else.

$\un{x}$ can then be mapped to an element in $W$ by $x = \zeta(\un{x})
= \di\sum_{j=1}^r\al_jg_{i_j}$ and a codeword $xu \in \C$ equated with
a codeword in $\cE$ given by $\zeta\I(xu)=\un{x}U$.

In summary, we obtain the $(n,r)$ code $\cE = \{\un{x}U: \un{x}\in
R^r\}$ in $R^n$ equivalent to $\C$. Considering codewords as $\un{x}U
\in \cE$ where $\un{x}\in R^n$ as described proves, as we will show,
convenient for analysis purposes.

For any $n\times n$ matrix $A$ let $\un{a}_1, \un{a}_2,
\ldots, \un{a}_n$ denote the rows of $A$ in order.

\paragraph{Generator for matrix-generated code} We can derive a
generator matrix $A$ for a code from $R^r$ to $R^n$ called the {\em
matrix-generated code}, and given by $\D=\{\un{x}A: x\in R^r\}$. It is
equivalent to codes $\C$ and $\cE$.

The codewords in $\cE$ consist of linear combinations of the
rows $\un{u}_{i_1}, \un{u}_{i_2}, \ldots, \un{u}_{i_r}$ of
$U$. Let $A$ be the $r \by n$ matrix consisting of the $i_1,
\ldots, i_r$ rows of $U$. We will show in \lemref{four} that the rows
of $A$ are linearly independent if and only if $Su$ is.

\subsubsection{Linear independence}
We now tie up the relationship between linear (in)dependent rows of
the $RG$-matrix $U$ and the linear (in)dependence of the set $Su$.

Suppose $S = \{g_{i_1},
g_{i_2}, \ldots, g_{i_r}\}$ inside $G= \{g_i,g_2, \ldots, g_n\}$ and
that $U$ is obtained from this listing of $G$. 
Specifically, it is established that the rows $\{\un{u}_{i_1},
\un{u}_{i_2}, \ldots, \un{u}_{i_r}\}$ of $U$ are  linearly independent if and
only if $Su$ are. 
\begin{theorem}\label{thm:dept}
Suppose $U$ has rank $t$. Let $S \subset G$ be a set of group elements
such that $|S|=t+1$. Then $Su$ is linearly dependent.
\end{theorem}
\begin{proof}
Let the rows of $U$ be $\un{u}_1, \un{u}_2, \ldots, \un{u}_n$ in
order.  

Suppose
$Su=\{g_{j_1}u, g_{j_2}u, \ldots, g_{j_{t+1}}u\}$.
Any $t+1$ rows of $U$ are dependent so there exists $\al_{j_1},
\al_{j_2}, \ldots, \al_{j_{t+1}}$ not all zero such that
$\di\sum_{k=1}^{t+1} \al_{j_k}\un{u}_{j_k} = 0_{1\times n}$. Let $A$
be the $RG$-matrix with first row having $\al_{j_k}$ in the $j_k^{th}$
position for $1\leq k \leq t+1$ and zeros elsewhere.

Then $A$ is the $RG$ matrix corresponding to the group ring element
$a= \al_{j_1}g_{j_1} + \al_{j_2}g_{j_2} + \ldots +
\al_{j_{t+1}}g_{i_{t+1}}$.  Also $AU$ is an $RG$-matrix whose first
row consists of zeros and hence $AU = 0_{n\times n}$.  Thus $au = 0$
and therefore $\{g_{i_1}u, g_{i_2}u, \ldots, g_{i_{r+1}}u\}$ is
linearly dependent as required.
\end{proof}

Then from the last
theorem it follows that we may take $S$ to have $r$ elements such $r
\leq \rank U$.  For if $r > \rank U$ then $Su$ is generated by $r$
elements $S'u$ (where $S' \subset S$), and the code is given by
$\C=W'u$ where $W'$ is the module generated by $S'$.

Alternatively, assuming $U$ has $\rank \geq r$, one can choose or find
$r$ linearly independent rows $\un{u}_{i_1}, \un{u}_{i_2}, \ldots,
\un{u}_{i_r}$ of $U$ and then construct $S$ by reference to these, let
\\ $S = \{g_{i_1}, g_{i_2}, \ldots, g_{i_r}\}$ and then $Su$ is linearly
independent.

Define $G_j$ to be
the $RG$-matrix corresponding to the group element $g_j \in G$ -- this
 is consistent with the notation for the $RG$-matrix corresponding to
 $g_j$. Then
 $G_j$ is the matrix whose first row has a $1$ in the $j^{th}$ position
 and zeros elsewhere. It is then
 clear that $G_jU$ is the $RG$-matrix with first row $\un{u}_j$. 

\begin{lemma}\label{lem:five} 
Suppose $\un{u}_1, \un{u}_2, \ldots,\un{u}_s$ are the first rows (or
first columns) of the $RG$-matrices $U_1, U_2, \ldots, U_s$
respectively. Then $\al_1\un{u}_1 + \al_2\un{u}_2 + \ldots +
\al_s\un{u}_s = 0$ if and only if $\al_1U_1+\al_2U_2 + \ldots
+\al_sU_s = 0$.
\end{lemma}
\begin{proof} Suppose   $\al_1\un{u}_1 + \al_2\un{u}_2 + \ldots +
\al_s\un{u}_s = 0$. Let $U = \al_1U_1+\al_2U_2 + \ldots
+\al_sU_s$. Then $U$ is an $RG$-matrix whose first row consists of
zeros and hence $U = 0$.

On the other hand it is clear that if $\al_1U_1+\al_2U_2 + \ldots
+\al_sU_s = 0$ then $\al_1\un{u}_1 + \al_2\un{u}_2 + \ldots +
\al_s\un{u}_s = 0$.
\end{proof}  

\begin{lemma}\label{lem:four} $\{ g_{i_1}u,g_{i_2}u, \ldots ,
g_{i_r}u\}$ is linearly independent if and only if \\$\{\un{u}_{i_1},
\un{u}_{i_2}, \ldots, \un{u}_{i_r}\}$ is linearly independent.
\end{lemma}
\begin{proof} Follows immediately from \lemref{five}.
\end{proof}. 

Suppose then $\rank U = r$ and that $\{\un{u}_{i_1}, \un{u}_{i_2},
\ldots, \un{u}_{i_r}\}$ is linearly independent.
Then by \lemref{four}, $Su$ is linearly independent. It is also clear
in this situation that $\C = RGu$, the right ideal generated by $u$.

\subsection{Check elements and matrices}
\slabel{checkmat} Throughout this section(\ref{sec:checkmat}), the
code under question is $(n,r)$ where $r=\rank U$. In
\sref{dim-less-than-rank}, we describe how to obtain check conditions
for $(n,k)$ codes where $k < \rank U$.

Clearly $cv = 0$ for any codeword $c$. The most convenient situation
is when the code $\C$ (and thus for codes $\D,\cE$) has a
(single) check element, i.e. that $y\in C$ is a codeword if and only if $yv=0$.
Equivalently, $V$ checks $\cE$ provided $\un{y}\in \cE$ if and only if
$\un{y}V=0$ if and only if $YV = 0$, where $\un{y}$ is the first row
of $Y$. We now examine these requirements for a check element.

\subsubsection{Check elements}
\begin{definition}
For a zero-divisor $u$ with $\rank U =r$, say $u$ is a {\em principal}
zero-divisor if and only if there exists a $v\in RG$ such that $uv =
0$ and $\rank V = n-r$.
\end{definition}

This is the situation for example when $RG$ is a principal ideal
domain as  when $G$ is a cyclic group --
see \sref{relationships} for a discussion and proof of this in the
cyclic group ring case. 

It is also possible in other cases that for a given zero-divisor $u$
there is a $v$ with $uv=0$ and $\rank U + \rank V = n$; for example if
$u^2 = 0$ or $uu^T = 0 $ and $\rank U = \frac{n}{2}$, in which case
$\rank U^T = \frac{n}{2}$ also.

Suppose that $uv = 0$ and $\rank V=n-r$. Then $y$ is a codeword if and
only if $yv = 0$ if and only if $YV=0$. This is not immediately
obvious and depends on the fact that $U$ and $V$ are $RG$-matrices;
the proof, in stages, is shown below.

\begin{lemma} Let $\un{y}$ be the first row of an $RG$-matrix
$Y$. Suppose also $V$ is an $RG$-matrix. Then $YV=0$ if and only if
$\un{y}V = 0$.
\end{lemma}
\begin{proof} Suppose $\un{y}V = 0$. Then $YV$ is an $RG$-matrix with
  first row consisting of zeros. Hence $YV = 0$. On the other hand if
  $YV = 0$ then clearly $\un{y}V = 0$.
\end{proof}

\begin{theorem}\label{thm:long} Let $\C=\{xu: x\in W\}$ where $W$ is
generated by $S$ such that $Su$ is linearly independent and $|S|=\rank
U=r$.  Suppose further that $uv=0$ in the group ring $RG$ so that
$\rank V=n-r$.  Then $y$ is a codeword if and only if $yv = 0$.
\end{theorem}
\begin{proof} (We need to show that $yv = 0$ if and only $ y= \al u$ for
some $\al \in W$.)  If $y$ is a codeword then $y = xu $ for some $x
\in W$ and hence $yv=0$.

Suppose on the other hand $yv=0$. Now $UV= 0$ where $U$ has rank $r$
and $V$ has rank $n-r$. The null-space of $V$ is the set of all (row)
vectors $\un{x}$ such that $\un{x}V = 0$. Since $V$ has rank $n-r$, by
linear algebra the rank of the null-space of $V$ is $r$. Since $U$ has
rank $r$ the rows of $U$ generate the null-space of $V$.

Since also $YV= 0$ the rows of $Y$ are in the null-space of $V$ and
hence the rows of $Y$ are linear combinations of the rows of $U$. In
particular $\un{y} = \un{a} U$ where $\un{y}$ is the first row of $Y$
and $\un{a}$ is a $1\times n$ vector. Let $Q$ be the $RG$-matrix whose
first row is $\un{a}$; $RG$ matrices are uniquely defined by their
first row. Then $QU$ is an $RG$-matrix whose first row is $\un{y}$,
the first row of $Y$. Hence $Y=QU$. From this it follows that $w = qu$
(where $q$ is the group ring element corresponding to the $RG$-matrix
$Q$). We need to show that $qu \in \C$. Let $q = \di\sum_{i=1}^n
\al_ig_i$. Suppose $g_j$ occurs in this sum with non-zero coefficient
and $g_j \not\in S$. Then $\{g_{i_1}u , g_{i_2}u, \ldots , g_{i_r}u,
g_ju\}$ is linearly dependent by \thmref{dept}, the first $r$ of which
is linearly independent. Hence $g_ju$ is in $\C$ as required.
\end{proof}

\begin{corollary} 
$\C=\{xu: x\in W\}$ has a single check element if and only if $u$ is a
principal zero-divisor.\endproof
\end{corollary}

\begin{corollary} $\un{y}\in \cE$ if and only if  $\un{y}V = 0$ if and only if $YV =0$ where $Y$ is the $RG$- matrix with first row $\un{y}$.\endproof
\end{corollary}

\subsubsection{General check conditions}
Define the null-space of $U$ to be $\Ker(U) = \{ \un{x} : U\un{x} =
0\}$ where $\un{x}$ is an $ n\times 1$ vector. Since $U$ has rank $r$,
the dimension of $\Ker(U)$ is $n-r$. Let $\un{v}_1, \un{v}_2, \ldots,
, \un{v}_{n-r}$ be a basis for $\Ker(U)$; these $\un{v}_i$ are
$n\times 1$ column vectors. Let $V_{i}$ be the $RG$-matrix with first
column $\un{v}_i$. Then clearly $UV_i = 0$ for $1\leq i \leq n-r$ since
$UV_i$ is the $RG$-matrix with first row consisting of zeros and hence
must be zero. Hence $uv_i = 0$ where $v_i$ is the group ring element
corresponding to the $RG$-matrix $V_i$.

Note that the null-space of $U$ is easily and quickly obtained using
linear operations on the rows of $U$. The basis for the null-space may
be read off from the {\em row-reduced echelon} form of $U$, which also
puts the generator in standard form.
This is also very useful in producing a check matrix for the
corresponding encoding $R^r \rightarrow R^n$ -- see \thmref{linear}
below.

Thus if $y$ is a codeword then $yv_i =0$ for $1\leq i \leq n-r$.
 The following theorem may be
proved along similar lines to \thmref{long}. Its proof is omitted.
\begin{theorem} Suppose $u$ is a zero-divisor, $\rank U= r$ and $W$ is
  generated by $S$ with $r$ elements such that $Su$ is linearly
  independent. Let $v_i$ be defined as above. Then $y\in \C$ if and
  only if $yv_i = 0$ for all $i=1,\dots,n-r$.\endproof
\end{theorem}

\begin{corollary} $Y$ is a codeword if and only if $YV_i = 0$
\end{corollary}

Not all the $v_i$ are needed -- just enough so that the corresponding
matrices $V_i$ contain a basis for the null-space.  In many cases a
particular $V_i$ of rank $n-r$ can be found.
The check conditions for the code $\cE$ follow:
\begin{theorem}\label{thm:linear} Let $\hat{V} = (v_1,  v_2, \ldots, v_{n-r}, 0, 0, \ldots , 0)$ be the $n\times n$ matrix with first $n-r$ columns consisting of $v_i$
in order and then $r$ columns with zeros.  Then $\un{y} \in \cE$ if
and only if $\un{y}\hat{V} = 0$.
\end{theorem}
\begin{proof} 
Clearly if $\un{y}\in \cE$ then $\un{y}\hat{V} = 0$.

Suppose then $\un{y}\hat{V} = 0$. Define $\Ker(\hat{V}) = \{t\in R^n :
t\hat{V} = 0\}$. Since $\hat{V}$ has dimension $n-r$, the dimension of
$\Ker(\hat{V})$ is $r$. Now each row of $U$ is in $\Ker(\hat{V})$
since $U\hat{V} = 0$. Since $U$ also has rank $r$ this implies the
rows of $U$ generate $\Ker(\hat{V})$. Hence $\un{y}$ is a linear
combination of the rows of $U$. The rows $u_{i_1}, u_{i_2}, \ldots,
u_{i_r}$ are linearly independent and hence are a basis for the row
space of $U$ which has dimension $r$. Hence $\un{y} = \al_1u_{i_1} +
\al_2u_{i_2} + \ldots + \al_ru_{i_r}$ and is thus a codeword.
\end{proof}

\subsubsection{Generator from a check element}
\slabel{checkelement-generator} The argument above may also be used to
obtain a generator when we use a zero-divisor $v \in RG$ to act as a
check element and produce the code $T_v= \{y \in:T yv=0\}$, regardless of
whether the code has a single generating element or not.

Take the case $T= RG$.
Suppose the resultant $RG$-matrix $V$ has rank $n-r$. Then $n-r$ of
the rows of $V$ are linearly independent and the other rows of $V$ are
linearly combinations of these. Thus the code may be considered a
$(n,r)$ code with check matrix of size $(n-r)\times n$.

Define $\Ker(V) = \{\un{x}: \un{x}V = 0\}$. Then $\Ker(V)$ has rank
$r$.  Let $\un{u}_1, \un{u}_2, \dots, \un{u}_r$ be a basis for
$\Ker(V)$. Form the matrix $\hat{U}$ with rows $\un{u}_i$. Then we get
the following:
\begin{theorem}
$\hat{U}$ is a generator matrix of the code $\cE$.
\end{theorem}
\begin{proof} The proof is similar to that of \thmref{linear}.
\end{proof}

\subsection{Check matrices when the dimension is less than the rank}
\slabel{dim-less-than-rank} We now take the case where $W$ is the
submodule generated by $S$ such that $|S|=s < r = \rank U$. This
generates an $(n,s)$ code.

One way to create a check matrix for $\D$ is to apply standard
row operations to obtain a basis for the null-space of the generator
$A$. However, it can also be obtained from the $RG$-matrices $U$ and
$V$ by adding certain $r-s$ vectors to $V$ as explained below.

Let $V_{n-r}$ denote a submatrix of $V$ consisting of $n-r$ linearly
independent columns.

Consider the indexing set $T=\{k_1, k_2, \ldots, k_s\}$ ($1 \leq k_1 <
k_2 < \ldots < k_s \leq n$) which defines
$S=\{g_{k_1},\dots,g_{k_s}\}$. Extend the set $T$ to a set of linearly
independent rows $R=\{k_1, k_2, \ldots, k_s, w_1 ,\ldots, w_{r-s}\}$
of $U$.

Let $U_r$ be the matrix formed from $R$ with the rows in order. Then
$U_r$ has rank $r$ and size $r\times n$. There exists an $n\times r$
matrix $C$ such that $U_rC = I_r$.

Delete the $k_1, k_2 , \ldots, k_s$ columns of $C$ to get an $n\times
(r-s)$ matrix, which we call $C_{r-s}$. We now add this $C_{n-r}$
matrix to $V_{n-r}$ to get the matrix $D$. This $D$ then has rank
$n-s$ and satisfies $U_rD = 0$. It follows that $\un{y} \in\D$ if and
only if $D\T \un{y}\T = 0$.

Thus $D\T$ is a check matrix for $\D$, obtained by adding certain
$r-s$ columns from $C$, the right inverse of $U_r$, to the matrix
$V_{n-r}$.

\subsection{Dual and self-check codes}
\slabel{zerodivisordual} By definition, a dual of a code $\C$
considered as vectors over $R^n$ is its orthogonal complement, namely
$\C^{\perp} = \{v \in R^n: \ip{v}{c}=0, \forall c \in \C \}$.

Let $x,y \in RG$. The inner (or dot) product, is given by term-by-term
multiplication of the elements in the ring $R$, namely $\ip{x}{y} =
\sum_{g\in G} \al_g \beta_g$ where $x = \sum_{g\in G} \al_g$ and $y =
\sum_{g\in G} \beta_g$.

Thus, the dual of a code from a group ring encoding is $\C^{\perp} =
\{y \in RG: \ip{ux}{y}=0, \forall x \in W \}$. We now show that the
dual of a zero-divisor code has an easy form.
\begin{theorem}
Let $u, v \in RG$ such that $uv=0$.  Let $U$ and $V$ be the
$RG$-matrices of $u$ and $v$ respectively, such that $\rank U=r$ and
$\rank V =n-r$.  Let $W$ be a submodule over a basis $S \subset G$ of
dimension $r$ such that $Su$ is linearly independent and $W^{\perp}$
denote the submodule over basis $G\less S$.  Then the code $\C = \{xu:
x \in W\}$ has dual code $\C^{\perp}=\{ x v\T: x \in W^{\perp}\} = \{
y \in RG: yu\T=0\}$.  \thmlabel{dualzero}
\end{theorem}
\begin{proof}
Note that $v\T$ is a zero-divisor and that $\rank V\T=n-r$ (because
$\rank V=n-r$), and that $W^{\perp}$ does not contain a zero-divisor
of $v\T$. Thus, there is a 1-1 map between $W^{\perp}$ and $\{ x v\T:
x \in W^{\perp}\}$.  It remains to show it is the dual.

Let $z\neq0$ be an element in $RG$. We need to prove that
$\ip{xu}{z}=0$, $\forall x \in W$ if and only if $z=y v\T$ for some $y
\in W^{\perp}$.

Suppose $z=y v\T$, and let $x,y \in RG$.

Recall that $\un{x}=\zeta\I(x),\un{y}=\zeta\I(y)$ are the vectors in
$R^n$ corresponding to $x,y$. Then, $\ip{xu}{z} = \ip{xu}{y v\T}=
\un{x}U (V\T \un{y})\T = \un{x}(UV)\un{y}\T = 0$.

Conversely, suppose $\ip{xu}{z}=0$, $\forall x \in W$. Without loss of
generality, assume $1\in W$. Then $\ip{u}{z}=0$ implies $z u\T = 0$,
and since $u\T$ is the check element for the code generated by $v\T$,
$z = y v\T$ for some $y \in W^{\perp}$.
\end{proof}

This is consistent with cyclic codes whereby the dual for a code with
generator $u$ and check $v$ is usually expressed as having the
reciprocal (polynomial) $g^{n-r}v\T$ (where the $\deg(v)=n-r$) as
generator~\cite{blahut03algebaic-codes}. Using this as generator with
$W$ having basis $\{1,g,\dots,g^r\}$ yields the same code as generator
$v\T$ with submodule $W^{\perp}$.

Of course, for the matrix-generated code $\D$, one may obtain a code
equivalent to its dual by interchanging the generator and check
matrices; see for example\cite{macwilliams_sloane}.

A condition for self-duality is an easy consequence.
\begin{corollary}
$\C^{\perp}=\C$ if and only if $u u\T=0$ and $\rank U =
n/2$. \endproof
\end{corollary}

A {\em self-dual zero divisor code} is thus a code given by $u$ with
$uu\T = 0$ and $\rank U = n/2$. We say a code given by $u$ is {\em
self-check} if $u^2=0$, in which case it is equivalently self-dual, as
the code and its dual are equivalent.

\subsubsection{Examples of zero-divisor self-dual codes}
\slabel{zd-selfdual}

Group rings are a rich source of elements $u$ such that $u^2 = 0$ or
$uu\T=0$ or both. These will generate self-dual (or self-check) codes
when $\rank U =n/2$. See \sref{selfdual1} for specific examples.

\section{Codes from units}
\slabel{codesfromunits}

In this section, we construct codes from units in group 
rings.
Let $u$ be a unit in the $RG$, where $G$ is of order $n$ and listed
$G= \{g_1, g_2, \ldots, g_{n}\}$. Let $W$ be a submodule of $RG$
generated (as an $R$-module) by  $r$ group elements 
$S=\{g_{k_1}, g_{k_2},\ldots,g_{k_r}\}$ with $r < n$.

As defined in \sref{encodingdefns}, the unit-derived code is 
$\C=\{ux: x \in W\}$ or $\C=\{xu : x \in W\}$. The code is thus constructed
from a unit $u$, a submodule $W$ and, when $RG$ does not commute, a choice
over left or right encoding.
Assume in what follows that the encoding is on the right ($x\mapsto xu$).
The left-encoding case $x\mapsto ux$ is similar, following, with minor
adjustments, the same general procedure.

Now $c$ is a codeword (i.e. in $\C$) if and only if $cu\I \in W$
i.e. {\em if and only if the coefficients of $G \less S$ in $cu\I$ are
zero}. Notice that multiplying a codeword by the inverse of the unit recovers
the original.

A unit-derived code can also be considered a mapping from $R^r$
to $R^n$. First, map a vector $\un{x} = (\al_1, \al_2, \ldots, \al_r)\in R^r$ by
$\lambda_{W}(\un{x})=\di\sum_{i=1}^r \al_i g_{k_i}$ to an element $x \in W$.
Then a codeword $xu \in \C$ is obtained which may be written $xu =
\di\sum_{i=1}^n \be_ig_i$. This gives an encoding $\un{x} \mapsto
(\be_1, \be_2, \ldots, \be_n)$ which is a map from $R^r\rightarrow
R^n$.

We also associate each unit-derived code with an equivalent code
which we call the {\em matrix-generated code} $\D$. This 
is a code from $R^r$ to $R^n$ and has an $r\by n$ 
generator matrix $A$ extracted from the $RG$-matrix $U$, and a check matrix 
that extracted from $V$. If $A$ is such a generating matrix, then 
$\D = \{\un{x}A:  \un{x}\in R^r\}$. The distinction between codes $\C$ and 
$\D$ is one of convenience. They are equivalent, exhibiting the same properties.
This procedure is conceptual and any practical implementation works
on producing $\D$ or $\C$ directly, depending on what is desired.

\subsection{Generator and Check Matrices}
Let us now examine the check and generator matrices that result from
a unit-derived code.
Suppose $u u\I = 1$ in the group ring and let $U,U\I$ respectively
be the corresponding $n\times n$ $RG$-matrices.

First, consider $W$ to be the submodule generated by $\{g_1, g_2, \ldots,
g_r\}$ with $r<n$ (i.e. has as basis the first $r$ elements in the chosen
listing of $G$). We later deal with the case when $W$ has a general basis of
group elements. An element in $W$ is thus of the form $x = \di\sum_{i=1}^r
\al_i g_i$. 

Divide $U = \begin{pmatrix} A \\ B \end{pmatrix}$ into block matrices where
$A$ is $r\times n$ and $B$ is $(n-r)\times n$. Similarly, let $U\I=
\begin{pmatrix} C & D\end{pmatrix}$ where $C$ is $n \times r$ and $D$ is
$n\times (n-r)$.

Now $AD = 0$ as $UU\I=I$. It is easy to see that $A$ is
a generator matrix for the matrix-generated code. We now show that $D\T$ is a
check matrix.
\begin{theorem}
Let $\un{y} \in R^n$ and $\D=\{\un{x}A: \un{x}\in R^r\}$.  Then
$\un{y} \in \D$  if
and only if $\un{y}D=0$.
\end{theorem}
\begin{proof}
If $\un{y}=xA$ for some $\un{x}\in R^r$, then clearly $\un{y}D=0$.
If, on the other hand, $\un{y}D=0$,
$$
\un{y} = \un{y}U\I U = \un{y}\begin{pmatrix} C & D \end{pmatrix}
 \begin{pmatrix} A \\ B\end{pmatrix} 
= \begin{pmatrix} \un{y}C & \un{y}D \end{pmatrix} \begin{pmatrix} A &
  B\end{pmatrix} = 
\begin{pmatrix}\un{y}C & 0\end{pmatrix} \begin{pmatrix} A \\
  B\end{pmatrix} =  \un{y} CA.
$$
Now $\un{y}C$ is in $R^r$ and $\un{y} = \un{y} U\I U = \un{x}A$ for
some  $\un{x}\in R^r$ as
required.
\end{proof}

Thus $D\T$ is a check matrix for the matrix-generated code $\D$: $\un{y}$
is a codeword if and only if $D\T \un{y}\T = 0$ if and only if $\un{y}D=0$.
The $r\times n$ generator matrix $A$ and $(n-r)\times n$ check $D\T$ produced
from this unit and submodule have full allowable rank, $r$ and $n-r$
respectively.

Units in group rings result in non-singular matrices, enabling the
construction of codes from units. Any non-singular matrix could also
produce a code by the above arguments, although of course
one could not exploit the underlying algebraic structure of a group ring.

When $W$ is generated by a general basis $S = \{g_{k_1}, g_{k_2}, \ldots,
g_{k_r}\}$, the generator and check matrices are obtained by `extracting' from
and `adding' to certain rows and columns from $U$ and $U\I$.
A generator matrix
results from the $r\times n$ matrix consisting of the $k_1, k_2,\ldots, k_r$
rows of $U$. Additionally, let $D$ be the $(n-r)\times n$ matrix obtained by
deleting the $k_1, k_2, \ldots, k_r$ columns of $V$. Then $D\T$ is a check
matrix.

\subsection{Constructing unit-derived codes}
The generator and check matrices for the matrix-generated code $\D$ are
immediate from the construction. However, working with the unit-derived
code $\C$ itself can be advantageous. For example, using the group ring check
conditions directly may be the best method for decoding.

In summary, unit-derived code of length $n$ and dimension $r$ can be
constructed quite freely as follows. Choose a group $G$
of order $n$ and a ring $R$ over which the code will be defined. Typically,
$R$ is a field but this is not a requirement; codes over the integers, rings
of matrices or other rings are also useful.

Find a unit $u$ in the group
ring $RG$ and its inverse $u\I$. 
As previously mentioned, if $R$ is a field or $RG$ of finite order,

every element in $RG$ is either a zero-divisor or a unit. When $R$ is a field,
there is a straightforward algorithm to determine which.
Generation of units is therefore not difficult.

Any basis for a submodule $W$ consisting of $r$ group elements will generate
a code, e.g. the first $r$ elements $\{g_1, g_2, \ldots, g_r\}$ according to
a listing of $G$.

It can proves advantageous to choose another, appropriate, basis, to say
increase the minimum distance of the code or optimise some other criteria.
This freedom
of basis nicely leads to the concept of an optimal one for a given unit $u
\in RG$ and dimension $r$ --- a so-called best-basis:
$$
\argmax_{S \subset G, |S|=r} \min_{x \in W(S)} \wt(xu)
$$
where $W(S)$ denotes the submodule generated by  $S$ and $\wt(y)$ the number of
nonzero coefficients of $y$.

This flexibility in choice of $r$ and $W$ and the full ranks obtained are major 
advantages of a unit-derived code over one derived from a zero-divisor.

\subsection{Derivation of Units} 
Group rings are a rich source of units. Units exist and are known in $RG$, where
$R$ can be any ring and not just a field, and from these, codes of different
types and makes can be constructed. Once a unit is known there is still a choice
on the submodule/dimension for the code and codes of different dimensions may be
obtained from a particular unit.

To fully describe a unit-derived code in terms of generator and check conditions
we need a unit and its inverse. The inverse may be known from the algebra;
general formulae for  certain units, and their inverses, in group rings are
known. Please consult \cite{milies_grouprings} and the 
references therein. In the cases 
of group rings over cyclic groups it is worth noting that the 
Euclidean Algorithm, which is extremely fast, may be used to obtain an 
inverse as then $RG \cong R[x]/\langle x^n-1\rangle$. A variation of 
the Euclidean
Algorithm  may also be used 
to find inverses in $RG$ when $G$ is a dihedral group.

Computer Algebra packages such as 
GAP, MAGMA which are particularly useful for handling groups, may in
addition  be used 
to find units in group rings;  for example, 
there exist function such as DihedralGroup() and DirectProduct() which return 
dihedral groups and direct product of groups respectively. 

The combination of units in a group ring is also a unit. This can be
exploited to produce new units which are not of the same form as the
originals and from which new codes can be derived.
Thus, for example,
bicyclic units could be combined with alternating units, Hoeschmann
units etc. to give new types of units. 
\subsection{Dual and Orthogonal Codes}
\slabel{unitdual}
Recall from \sref{zerodivisordual} that the dual of a code from a group ring
encoding is $\C^{\perp} = \{y \in RG: \ip{ux}{y}=0, \forall x \in W \}$ and 
the concept of the transpose of a group ring element (\dref{transpose}). We
now show that the dual of a unit-derived code can be generated from $(u\I)\T$.
\begin{theorem}
Let $W$ be a submodule with basis of group elements $S \subset G$ and
$W^{\perp}$ be the submodule with basis $G\less S$. Let $u\in RG$ be a unit
such that $u u\I=1$. Then the dual code of $\C = \{xu: x \in W\}$
is $\C^{\perp} = \{ x (u\I)\T: x \in W^{\perp}\}$.
\thmlabel{dualunit}
\end{theorem}
\begin{proof}
Let $z\neq0$ be an element in $RG$. We need to show that
$\ip{xu}{z}=0$, $\forall x \in W$ if and only if $zu\T \in W^{\perp}$.

Note that $\ip{xu}{y (u\I)\T} = \ip{x}{y}$. Thus, if $zu\T \in W^{\perp}$,
then, for all $x\in W$, $\ip{xu}{z} = \ip{x}{zu\T}=0$.

Conversely, if $zu\T \in W$, pick a $g\in S$ that has a non-zero coefficient
$\gamma$ in $zu\T$. Then, $\ip{gu}{z} = \ip{g}{zu\T} = \gamma \neq 0$.
\end{proof}

Strict equivalence of a unit-derived code and its dual, whereby
$\C=\C^{\perp}$, requires that for all $x
\in W$, $xuu\T \in W^{\perp}$, which imposes an impractical restriction.
However, it is natural to say that a unit-derived code is {\em self-dual}
if $\C$ and $\C^{\perp}$ are equivalent codes, or equivalently, that the
resultant matrix-generated codes $\D$ and $\D^{\perp}$ are equal.
\begin{definition}
An unit $u\in RG$ is {\em orthogonal} if and only if its inverse is $u\T$
(i.e. $uu\T = 1$).
\end{definition}

It is easy to see that the $RG$-matrix from an orthogonal unit $u$ is an
orthogonal matrix. From the above, it can also be seen that an
orthogonal unit combined with a submodule of dimension $n/2$ generates a
self-dual unit-derived code.

\section{Examples codes}
\label{sec:examplecodes}
In this section, we explore some illustrative constructions of
codes from group ring encodings. 
The examples below include codes in general abelian
groups and also in non-abelian groups. 

\subsection{Self-dual codes}\slabel{selfdual1}
Some self-dual codes in $RG$ can be formed as follows. Suppose $|G| =n= 2m$ and $ G = \{g_1, g_2, \ldots , g_n\}$. Let
$u \in RG$ satisfy:
\begin{enumerate}
\item $u^2 = 0$.
\item $u = u\T$ so that $uu\T = 0$.
\item $u$ and its corresponding matrix $U$ have $\rank m$.
\end{enumerate}

Then $u$ generates a self-dual code. Here's a specific example.
Let $G= C_2\times C_4$ where $C_4$ is generated by $a$ and $C_2$
 is generated by $h$. Form the group ring $\Z_2G$.

Consider $u = 1 + h(a + a^2 +a^3)$. Then $u^2 = 1 + h^2(a^2 + a^4 +
a^6)= 1 + a^2 + 1 + a^2 = 0$. Thus $\rank u \leq 4$. The $RG$-matrix of $u$ is
$U = \left(\begin{array}{rr} I & B \\ B & I\end{array}\right)$ 
from which it follows that $\rank u = 4$. By algebraic techniques on the group ring it can be shown that the distance is 4.
We thus get a $(8,4,4)$ self-dual code -- this must then be the extended Hamming self-dual code.

Extending this by considering $G = C_4^n \times C_2$ or other direct products is the subject of further work and produces an infinite series of self-dual codes with increasing distance.

\subsubsection{Example of unit-derived  codes} 
\slabel{unit-selfdual}

If $u^2 = 0$ then  $(1+u)^2 = 1$ over $\Z_2$. 
Consider $u_i = g^i+g^{n-i}+g^{n+i}+g^{2n-i}$  in $\Z_2C_{2n}$, with $C_{2n}$
  generated by $g$. Then $u_i^2 = u_iu_i\T = 0$ and thus any
  combination, $u$ say, of the $u_i$ satisfies $u^2 = uu\T
  =0$. Consequently $a = 1 +u$ satisfies $a^2 = aa\T = 1$, and gives a
  series of orthogonal units. 
There is no problem with the rank as we are dealing with unit-derived codes.

A specific example of this is as follows: In $\Z_2C_{14}$, 
$u = 1 +g^2+g^5+g^9 + g^{12}$ satisfies $u^2 =
uu\T=1$. 
The code has distance $d =
4$, and thus we get a $(14,7,4)$  code, with the best
possible distance for a $(14,7)$ binary code.

\subsection{Dihedral codes}
\label{sec:dihedral}
The first natural series of
non-abelian groups are the dihedral groups.

Codes from group ring encodings from the dihedral group appear to offer great
potential. There has been prior investigation into codes from ideals in the
dihedral group algebra as early as 1969\cite{macwilliams69-codesfromideals}, 
and a more recent result~\cite{bazzi03-groupactions}, which showed that there
exist a random ideal in $Z_2 D_{2n}$ for infinitely many $n$, such that the
resultant code of rate $1/2$ is ``good".

The focus on using encodings directly and the use of matrix algebra $RG$-matrix
structure is, to the best of our knowledge, a new approach.

The dihedral group $D_{2n}$ of order $2n$ is given $D_{2n} = \langle
a, b : a^2,  b^n, ab = b^{-1}a\rangle$. There are a number of possible
listings of the elements of $D_{2n}$, of which 
$D_{2n} = \{1, b, b^2, \ldots , b^{n-1}, a, ab, ab^2 ,
\ldots , ab^{n-1}\}$ proves most convenient.

An element $u \in R D_{2n}$ can be written
$$
u=\di\sum_{i=0}^{n-1} \al_i b^i + \di\sum_{i=0}^{n-1} \be_i ab^i.
$$
The associated $R D_{2n}$-matrix $W$ is then,
\newcommand{\AMAT}{\begin{matrix}
\al_0 & \al_1 & \al_2 & \ldots &\al_{n-1} \\
\al_{n-1} & \al_0 & \al_1& \ldots &\al_{n-2} \\
\vdots & \vdots & \vdots & \vdots & \vdots \\
\al_1 & \al_2 & \al_3 & \ldots & \al_0
\end{matrix}}
\newcommand{\BMAT}{\begin{matrix}
\be_0 & \be_1 & \be_2 & \ldots & \be_{n-1} \\
\be_1 & \be_2 & \be_3 & \ldots & \be_0 \\
\vdots & \vdots & \vdots & \vdots & \vdots \\
\be_{n-1} & \be_0 & \be_1 & \ldots & \be_{n-2}
\end{matrix}}
$$
U=\left(\begin{array}{l|l} 
\AMAT & \BMAT \\
\hline
\BMAT & \AMAT
\end{array}\right).
$$
This can be written $U = \begin{pmatrix} A & B \\ B & A \end{pmatrix}$,
where $A$ is circulant. Now, $B$ is a reverse circulant matrix as 
each row is a  circulant shift to the left of the one previous. Interestingly,
in a non-group ring 
context, reverse circulants have appeared before in 
codes~\cite{huffmann03fundamentals_ecc}.

A useful method for classification of the units and zero-divisors in
such a group ring results. Suppose $R$ is an integral domain not of
characteristic 2. Then, in general, $R D_{2n}$ is isomorphic to the
ring of matrices of this form. This ring of matrices is, in turn,
isomorphic to the ring of matrices of the form $\begin{smatrix} A + B
  & 0 \\ 0 & A - B \end{smatrix}$ with $A,  B$ as before. These
results are shown in \cite{hurley_grouprings2006}.

A matrix of this form is invertible, in an integral domain of 
characteristic not 2,  if and only if $A+B$ and $A-B$ are
invertible. Thus, $u \in R D_{2n}$ is a unit if and only if $A+B$ and
$A-B$ are. If $R$ is a field, then $u$ is zero-divisor if it is not a
unit (\thmref{groupalgebra}). Otherwise, it may similarly be
determined if $u$ is a zero-divisor.

Any multiplication $xy$ in $R D_{2n}$ or, equivalently, using the
corresponding $RG$-matrices can be done with low-complexity using 
existing techniques on circulant matrices (such as Fast Fourier
Transform methods). $A$ is already circulant, and the operation $xB$
can be done by calculating $xB'$ where $B'$ is the ``flip" of $B$.

\subsubsection{Dihedral zero-divisor codes}
\slabel{dihedral-zd}
For a zero-divisor $u \in R D_{2n}$ where $R$ is a field, finding a basis $S$
for the submodule $W$ can be done by a simple algorithm. Pick, in order,
elements from the set $\{1, b, b^2, \ldots , b^{n-1} \}$ until the first $k$
such that $\{u, bu, b^2u, \ldots , b^ku \}$ is linearly dependent, or else they
are all linearly independent.  Then add elements, in order, from $\{a, ab, ab^2,
\ldots , ab^{n-1} \}$ until the combined set $\{u, bu, \ldots , b^{k-1}u, au,
abu, \ldots, ab^l\}$ is linearly dependent. This is shown, by
\thmref{dihedral-independence}, to give maximum rank. An equivalent process
can be performed using the $RG$-matrix $U$ directly.

Consider the following code of 
rate $1/2$, constructed from a zero-divisor in $\Z_2D_{2n}$ for $n$ even. Let $u
=
1 + a + ab + \cdots + ab^{n-2}$. Then  $v= b + b^2 + \cdots + b^{n-1} +
ab^{n-1}$ satisfies $uv=0$. 
The $RG$-matrix of $u$ has the form $U=\begin{pmatrix} I_n & B\\ B & I_n
\end{pmatrix}$ where 
$B= \begin{ssmatrix} 1 & 1
& \ldots &1 & 0 \\ 1 & 1 & \ldots & 0 & 1 \\ \vdots & \vdots &\vdots
&\vdots & \vdots \\ 0 & 1 & \ldots & 1 & 1 \end{ssmatrix}$
is a reverse circulant matrix, consisting of all ones in first row except for a zero 
in the last entry, with
subsequent rows determined by the first. It is easy to see that $\rank U=n$. 

Now, $V=\begin{pmatrix} E & F \end{pmatrix}$. Here, $\rank V = n$. As $\rank U
+ \rank V = 2n$, $v$ is a check element for the code when the submodule has
dimension $n$, and the code generated is an ideal in $\Z_2 D_{2n}$.

This code has length $2n$ and dimension $n$, and thus rate $1/2$.
The encoding $\Z_2^n \rightarrow \Z_2^{2n}$ has a generator and a check
matrix $\begin{pmatrix} I_n & B\end{pmatrix}$ and $E\T$ respectively. A generator
matrix of the form $\begin{pmatrix} I_n & B\end{pmatrix}$, where $B$ is a
reverse circulant matrix, is known as reverse circulant generator 
matrix~\cite{huffmann03fundamentals_ecc}. It is thus clear that codes with
these generator matrices always arise from elements in $RD_{2n}$ of the form
$1+au$, where $u$ can be considered an element from $RC_n$.

Now consider the following example of codes over $D_{2n}$ built up from elements
in $C_n$. Let $u\in RC_n$ be a zero-divisor such that $uv=0$ where $C_n$ has
generating element $b$. From it we can construct a zero-divisor of the form $u +
axu \in D_{2n}$ for any $x\in RC_n$ with $(u + axu)(v+ayv\T)=0$ for any $y\in
RC_n$. For simplicity, we restrict ourselves to the case $x=y=1$. Additionally,
any element $u\in \Z_2 C_n$
(including units) will produce a zero-divisor in $\Z_2 D_{2n}$ of the form
$u+au$.

This simple construction can produce surprisingly decent codes.
Consider $u=1+b^2+b^5 \in \Z_2 C_7$, which produces the Hamming (7,4,3) code.
The zero-divisor $u+au$ has a corresponding $RG$-matrix with rank $7$ and
produces a $(14,7,4)$ code which is the best possible for this length and
dimension. Similarly $u=1+b^2+b^3+b^9+b^{10}+b^{11}$ produces a $(24,11,8)$
code -- also the best distance possible for $(24,11)$ binary code.
For $u \in \Z_2 C_{31}$ given by
$1+g+g^6+g^9+g^{10}+g^{14}+g^{15}+g^{16}+g^{17}+g^{19}+g^{20}+g^{21}+g^{22}+g^{
23}+g^{25}+g^{27}$ yields a $(62,30,12)$ code over $\Z_2 D_{62}$ which has the
same distance as the best-known $(62,30)$ code (it could be possible codes
with better distance may exist).

\subsection{LDPC codes}\label{sec:ldpc2}
In general, a code is an LDPC code if its check matrix is sparse, with few
non-zero entries. Thus, a code from a group ring encoding will be LDPC if 
the check element is short, namely contains few terms. LDPC unit-derived
codes are obtained by finding a unit element $u \in RG$ so that either $u$ or
$u^{-1}$ has only a small number of non-zero coefficients relative to the size
of the group. It is then possible to describe whole series of such codes from
group rings. 

Although most LDPC codes have been produced by randomised techniques, there has
been recent activity in the area of
algebraic constructions~\cite{tang04-algebraicldpc,
milenkovic04blockcirculantldpc, tanner2004-qcldpc}. The group ring encoding
framework is a useful tool in this direction, as it seems (as we shall show)
that some of  these constructions are implicitly working in a group ring.  
Algebraic constructions have some advantages, including potentially the ability
to have immediate generator matrices (it often proves difficult to find one from
a randomly produced check matrix) and ascertain performance without extensive
testing. One could also envisage a hybrid whereby a random construction is
performed within the parameters of an algebraic construction.

\subsubsection{Unit-derived Example}
\label{sec:ldpc3}
LDPC unit-derived codes with no short cycles can be achieved. These are neither 
cyclic nor an ideal. Consider for example the unit-derived code with check element 
$v=1 + g^{n-1} + g^{n-3} + g^{n-8} + g^{n -12}$ in $\Z_2C_{1000}$. 
We omit the generator element $u$
as it has 481 non-zero elements (it is quickly calculated using the Extended 
Euclidean Algorithm). 

Let $W$ be the module generated by $\{1,g^2, g^4, \ldots,
\}$ in $\Z_2C_{1000}$. Then the generator matrix for the code is the
matrix obtained by taking its first, third, etc. rows from the $RG$-matrix
of $u$ and the check matrix, which will be sparse, is obtained by deleting the
first, third etc. columns of the $RG$-matrix of the check element $v$ and then 
transposing.
\subsubsection{Zero-divisor Examples}
The regular $(j,k)$ LDPC codes from~\cite{tanner2004-qcldpc}, called
quasi-cyclic (QC) LDPC codes, are algebraically constructed. Interestingly,
the code can be obtained by shortening a code from an encoding in the group ring
$\Z_2G$ where $G=C_m \times C_k \times C_j$ i.e. the resultant parity-check
matrix for code of block length $mk$ can be obtained from say $v \in RG$. For
simplicity, take the 
case $j=2$, $k=3$, (a generalisation is straightforward). Let $C_3$ and $C_2$
have generators $a$ and $b$ respectively. Noting that $\Z_2G \isomorphic
(\Z_2C_m) (C_3 \times C_2$) 
one can write the
$RG$-matrix in this group ring:
$$
V = \begin{pmatrix}
I_{1} & I_{a} & I_{a^2} & I_b & I_{ab} & I_{a^2b}\\
I_{a^2}& I_1 & I_a & I_{a^2} & I_{b} & I_{ab}\\
I_{a} & I_{a^2} & I_1 & I_{ab} & I_{a^2b} & I_{b}\\
I_{b} & I_{ab} & I_{a^2b} & I_1 & I_a & I_{a^2}\\
I_{a^2b} & I_b & I_{ab} & I_{a^2} & I_1 & I_a\\
I_{ab} & I_{a^2b} & I_b & I_a & I_{a^2} & I_1
\end{pmatrix}
$$
where each entry is a $m \times m$ circulant permutation matrix determined in
the same fashion as~\cite{tanner2004-qcldpc}. Taking the first $3$ elements of
every 3rd row gives the parity-check matrix used in~\cite{tanner2004-qcldpc}. In
general, the parity-check matrix will be the same except for row ordering.

As the authors state, the resultant parity-check matrix may not have all
linearly independent rows, resulting in a code which may exceed the target rate
$1-j/k$ (although only slightly as they discover). Applying knowledge of the
group ring may help in achieving the target rate.

The LDPC construction in~\cite{milenkovic04blockcirculantldpc} can also be seen
as construction from a group ring, in this case $\Z_2 (C_N \by C_S)$, resulting
in a parity-check matrix from the 1st $mN$ rows of the $RG$-matrix, and each
block $N\by N$ matrix is a circulant permutation matrix.

Inspired by the above constructions, we propose a regular $(mk,j,k)$ LDPC code
with target rate $1-j/k$, from a group ring $R (G \cross H)$, where
$G=\{g_1,g_2,\ldots,g_m\}$ and $H=\{h_1,h_2,\ldots,h_k\}$ are groups of order
$m$ and $k$ respectively ($m>k>j$).

This can be from either a zero-divisor or unit code. The $RG$-matrix from 
$R (G \cross H)$ group ring is a $k \times k$ block matrix where each block is an 
$m \cross m$ $RG$-matrix in $RG$.

Recall that $G \cross H= \{(g,h): g\in G, h\in H\}$. Over $Z_2$, construction
amounts to deciding which group elements to choose from all possible ones.
The general idea is to construct an element in the group ring of weight $k$ by
picking, for each $h \in H$, exactly one element in $f(h) \in G$, yielding
element $v = \sum_{h\in H} (f(h),h)$ ($f:H\to G$ is not necessarily $1-1$ so it is
possible that $f(h_i)=f(h_j)$ for $i\neq j$). The subsequent $RG$-matrix is,
$$
V = \begin{pmatrix}
I_{h_1} & I_{h_2} & \dots & \dots & I_{h_k}\\
I_{h_2\I h_1} & I_{h_1} & \hdotsfor{2} & I_{h_2\I h_k}\\
\hdotsfor[2]{5}\\
I_{h_k\I} & I_{h_k\I h_2} &\hdotsfor{2} & I_{h_1}
\end{pmatrix}.
$$
Each block matrix $I_{h_l}$ has entries $\alpha_{st} = 1$ if $g_s\I g_t=f(h_l)$
and zero otherwise. $V$ thus has exactly $k$ ones in each row and column. 

The submodule $W$ for this code is chosen to have dimension $mj$; $j$
rows of block matrices are picked from $V$, resulting in a parity
check matrix $C$.  Since each $I_{h_l}$ has exactly one $1$ in each
column, $C$ will have $k$ ones in each column and $j$ ones in each
row, and is thus a check for a regular $(j,k)$ LDPC code.

For a given $k$ and $m$ there are many possible $j$. An exact rate
$1-j/k$ can be achieved when $v$ is a unit. When $v$ is a
zero-divisor, then it can be attained whenever a judicious choice of
basis $S \subset G\times H$ is possible such that $Sv$ is linearly
independent (see \sref{choosing-basis}).

Initial results are promising, including, over $\Z_2 (C_k \by C_m)$
for $j=3$ and $k=4$, $(31,26,3)$ and $(21,5,10)$ binary codes which
have the best possible distance possible in their class.

\section{Some Relationships with known codes}
\label{sec:relationships}

\subsection{Cyclic codes}
We show here that cyclic codes are exactly zero-divisor codes in group
rings on cyclic groups for special cases of the module $W$.  Also noteworthy 
is that Reed-Muller codes are extended cyclic codes and have been shown to be
associated with the group ring of the elementary abelian
2-group~\cite{assmus_reedmuller}.

For a given polynomial $h(g)$ over a ring $R$, let $r(g)$, be the
polynomial of minimal degree such that $h(g)r(g) \equiv 0 \mod (g^n -
1)$ (whenever it exists). Then, a cyclic code over $R$ is generated by
$h(g)$ with corresponding check polynomial $r(g)$.

The group ring $RC_n$ of the cyclic group $C_n$ over
a ring $R$ is isomorphic to $R[g]/\langle g^n-1\rangle$. This is a 
well-known result (e.g. ~\cite{blahut03algebaic-codes}).
Cyclic codes of degree $n$ are given by 
zero-divisors in $RC_n$ and the check
matrices the counterpart of the zero-divisor.
\begin{theorem}
Let $h(g) = \al_01 + \al_1g + \ldots + \al_rg^r$.

$h(g)$ is the generator polynomial of a cyclic code of length $n$ if
and only if the $RG$-matrix \\
$M(RC_n, h(g)) = \sigma(h(g))$ is a zero divisor in $R_{n\times n}$.

The check matrix of the code is given by the polynomial or group ring
element $r(g)$ such that $h(g)r(g) \equiv 0 \mod (g^n - 1)$.
\end{theorem}
\begin{proof}
If $h(g)$ is a zero-divisor in $R(G)$ then $h(g)r(g) = l(g)(g^n-1)$ as
polynomials.

The code generated by $h(g)$ is the same as the one generated by \\$d(g)
= \gcd(h(g),g^n-1)$ and then $d(g) p(g) = g^n -1$.

$M(RG,d(g))$ gives the generator matrix of the code and $M(RG, p(g))$
the check matrix.
\end{proof}

Cyclic codes in standard notation are presented using non-square
matrices. However, the matrices can be made square by considering them
as group ring elements.

In cyclic codes, the matrix representation of $u$ is used and the
mapping is given as $x \mapsto xu$ where $x \in R^r$ and $u\in
R_{r\times n}$. Consider $x$ as an element in $R^n$ (by adding 0's)
and the element $u$ as an element in $R_{n\times n}$ which is the
group ring completion of the matrix $u$. Then the mapping is $x
\mapsto xu$ in matrix form and is equivalent to the previous mapping.

See also \sref{ideals} on ideals in the cyclic group ring.

\subsection{Complexity relationship to standard mappings}
Codes from group ring encodings are not complex to implement. Since the
first row of an $RG$-matrix specifies the entire matrix, any operations on them 
can be quickly performed.

Normally, codes are considered maps $\beta: R^r \rightarrow R^n$ with $r<n$. 
The codes  from group ring encodings are $x \mapsto xu$ and these have equivalent 
encodings in matrix form $R_{n\times n} \rightarrow R_{n\times n}$
given by $\al: X \mapsto XU$. Now $X$ is an $RG$-matrix with $0$ in
$n-r$ entries of its first row and $X$ and $XU$ are  determined by
their first rows. Thus the mappings $\beta$ and $\al$ require the same number of
calculations.

\subsection{Ideals}
\label{sec:ideals} 

Here we discuss conditions under which codes as defined in
\sref{gr_codes} are ideals. It transpires that unit-derived codes as
defined in \sref{codesfromunits} are never ideals and that zero-divisor codes
as defined in \sref{zerodivisors} are ideals only in very special cases.

\begin{definition}
$I$ is said to be an {\em ideal} in the
ring $H$  if $I$ is a subring
such that (i) $hi \in I$, $\forall h \in H , \forall i \in I $ and (ii)
$ih \in I$,  $ \forall h \in H, i \in I$. Such an ideal is often
referred to as a {\em two-sided} ideal. $I$ is said to be a {\em left
ideal} in $H$ if $I$ is a subring such that $hi \in I, \forall i \in
I, \forall h \in H$. $I$ is said to be a {\em right ideal} in $H$ if
$I$ is a subring of $H$ such that $ih \in I, \forall i \in I, \forall
h \in H$. 
\end{definition}

In the case of a commutative ring there is no distinction
between left, right and two-sided ideals.

The cyclic codes are ideals in the group ring on the cyclic group 
but for example the
quasi-cyclic and shortened cyclic codes are not ideals. Our intention
here is to clarify the situation with respect to codes in group rings
as defined in \sref{gr_codes}.

Recall that such a code is either $Wu$ or $uW$ where
$W$ is a submodule of $RG$ and $u \in RG$. We consider the right encoding
$Wu$; the other one is similar.
Assume that $G= \{g_1, g_2, \ldots, g_n\}$ and 
here consider cases where the ring $R$ is a field.

Assume also that $W$ is generated by 
$S = \{g_{i_1}, g_{i_2}, \ldots , g_{i_r}\}$. The
code is then $\C = \{xu: x \in W\}$ and is  
generated by $Su = \{g_{i_1}u, g_{i_2}u, \ldots, g_{i_r}u\}$.  
We may assume as explained in \sref{zerodivisors} that $Su = \{g_{i_1}u,
g_{i_2}u,
\ldots, g_{i_r}u\}$ is linearly independent.

The  rows of an $n\times n$ matrix $U$ are designated in order by $\un{u}_1,
\un{u}_2, \ldots, \un{u}_n$.

Define $G_j$ to be the
$RG$-matrix corresponding to the group element $g_j \in G$. Thus $G_j$
is the matrix whose first row has a $1$ in the $j^{th}$ position and
zeros elsewhere and this first row determines the $RG$-matrix
$G_j$.  It 
then follows $G_j U$ is the $RG$-matrix with first row $\un{u}_j$. 
\begin{lemma}\label{lem:two}
Suppose $g_{i_1}u, g_{i_2}u, \ldots, g_{i_r}u$  is
  linearly independent. Then $\rank U \geq r$.
\end{lemma}
\begin{proof} Since $\{g_{i_1}u, g_{i_2}u, \ldots, g_{i_r}u\}$  is
  linearly independent so is \\ $\{G_{i_1}U, G_{i_2}U, \ldots,
  G_{i_r}U\}$. Thus $\un{u}_{i_1}, \un{u}_{i_2}, \ldots, \un{u}_{i_r}$
  is  linearly
  independent and hence $U$ contains $r$ linearly independent rows.
\end{proof}

\begin{theorem} Let $\C$ be the code $Wu$ with $W$  generated by $
  S= \{g_{i_1}, g_{i_2}, \ldots, g_{i_r}\}$ 
such that $Su =  \{g_{i_1}u,
  g_{i_2}u, \ldots, g_{i_r}u\}$ is linearly independent. Then $\C$ is a
  left  ideal if and only if $\rank U = r$.
\end{theorem}
\begin{proof} Suppose $\rank U = r$. We wish to show that $Wu$ is an
  ideal in $RG$. It will be sufficient to show
that $gu$ is in $\C$ for any $g \in G$. 

Since $g_{i_1}u,  g_{i_2}u, \ldots, g_{i_r}u$ is linearly independent
then by
\lemref{four} of \sref{zerodivisors}
  $G_{i_1}U, G_{i_2}U, \ldots, G_{i_r}U$ is linearly independent and
  so the rows $\un{u}_{i_1}, \un{u}_{i_2}, \ldots, \un{u}_{i_r}$ of
  $U$ are  linearly
  independent. Since $U$ has rank $r$ these rows thus form a basis for
  the row space of $U$.

Suppose then $g\in G$. If $g\in S$ then clearly $gu \in
\C$.

Suppose $g \not\in S$. Then $g= g_j$ where $j$ is not any of $i_1,
i_2, \ldots, i_r$. Now $G_jU$ is the $RG$-matrix whose first row is
$\un{u}_j$, the $j^{th}$ row of $U$.

Then $\un{u}_j = \di\sum_{k=1}^{r} \al_ku_{i_k}$ for some $\al_k \in R$
since the rows in the sum are a basis for the row space of $U$. Then
by \lemref{four} of \sref{zerodivisors} it follows that 
$G_jU = \di\sum_{k=1}^{r} \al_k G_{i_k}U$.
Hence
$g_ju = \di\sum_{k=1}^{r}\al_kg_{i_k}u$ and thus $g_ju \in \C$ as required.

Suppose on the other hand $\C$ is a left ideal. Then $g_iu \in\C$ for the 
$i^{th}$ element $g_i \in G$. Then as before $\un{u}_j$ is a linear
combination of $\un{u}_{i_1}, \un{u}_{i_2}, \ldots ,
\un{u}_{i_r}$. Hence $\rank U\leq r$. Since $\un{u}_{i_1},
\un{u}_{i_2},  \ldots ,
\un{u}_{i_r}$ is linearly independent it follows that $\rank U = r$. 
\end{proof} 

Note that the proof also shows that when $\C$ is a left ideal, then
$\C = Wu = RGu$.

If $u$ is a unit then $\rank U=n$. Hence $\C$ is not an ideal in a 
unit-derived code - the module $W$ is never equal to all of $RG$ in our
definition of unit-derived code.

Let $I$ be a left ideal in a ring $R$. Similar remarks apply to right
and two-sided ideals. If $I$ contains a unit then clearly
$I$ contains the identity and hence $I = R$. Now $I$ is said to be a
{\em proper} ideal if $I\not = 0$ and $I \not = R$. Here again we can see
that a unit-derived code is never an ideal as this has the form $Wu$
where $u$ is a unit  and thus $Wu$ has
a unit.

Suppose $I$ is a proper left ideal of a group ring $RG$ where $R$ is a
field. Then $I$ is a subspace of $RG$ and so is generated as an 
$R$-module by $u_1, u_2, \ldots,
u_s$ with $u_i\in RG$. Since $I$ is a proper ideal none of the $u_i$
can be  units and
so by \thmref{groupalgebra} each $u_i$ is a zero-divisor. Also no
linear combination of
the $u_i$ can be a unit as $I$ does not contain any unit. Thus:
\begin{theorem} $I$ is a proper ideal in a group ring $RG$,
  with  $R$  a field, if and only if $I$ is generated as a module by  $u_1,
u_2, \ldots, u_s$ where each $u_i$ is a zero-divisor and no linear
combination of the $u_i$ is a unit. 
\end{theorem}

A left ideal $I$ is {\em principal} in $RG$ if and only if it has the form
$RGu$  for an
element $u\in RG$. Thus proper principal ideals in the group ring over
a field are of the form $RGu$ where $u$ is a zero-divisor. These are the
particular zero-divisor codes where the rank of $U$ equals the number of
elements in the generating set $S$ of $W$ where $Su$ is 
linearly independent.
\subsubsection{Ideals in cyclic group rings}
\label{sec:cyclicideal}
Because of its particular nature as a polynomial type ring, it is easy
to show that all ideals in
the cyclic group ring are principal.
\begin{lemma} Every ideal in the group ring of the cyclic group is
  principal.
\end{lemma}
\begin{proof}  Let $I$ be an ideal in the group ring of the cyclic
  group $RG$ where $G$ is generated by $g$. Choose $f(g)$ in $I$ of
  minimal degree in $g$. Let $x(g) \in I$. Then by Division Algorithm
  to polynomials, $x(g) =  q(g)f(g) + r(g)$ where $r(g) = 0 $ or $r(g)$
  has degree less than $f(g)$. Now $r(g) \in I$ since $x(g)\in I$ and
  $f(g) \in I$. Since $f(g)$ is of minimal degree in $I$ this implies
  $r(g) = 0$ and $x(g) = q(g)f(g)$. Hence $I$ is a principal ideal
  generated by $f(g)$.
\end{proof}

Suppose now we have a zero-divisor $u$ in the cyclic group ring $RG$
and that $\rank U = r$. In order for $u$ to be a principal zero-divisor
we require an element $v \in RG$ such
that $uv=0$ and $\rank V =n-r$. Such a $v$ always exists in the
cyclic group ring. Choose $v$
to be an element of least degree such that $uv = 0$. Here in fact what
we are doing is choosing a generator for the annihilator of $u$  which
is a principal ideal.
\begin{theorem}Suppose $u$ is a zero-divisor in the cyclic group ring
$RG$ with $G$ of order $n$. Let $v$ be an element of least degree
such  that $uv=0$. If $\rank U =r$ then $\rank V = n-r$.
\end{theorem}
\begin{proof} If $uv_1 = 0$ then by Division Algorithm, $v_1 = qv +
  r$ where $r=0$ or $\deg(r) < \deg(v)$.
Then multiplying through by $u$, and noting that elements commute, we see
that $ur = 0$. Since $v$ is of least degree such that $uv =0$ this implies that $r=0$ and hence $v_1 = qv$.

We now show that the null-space of $U$ is generated by the rows of
$V$. The null-space of $U$ is the set of all vectors $\un{x}$ such that
$U\un{x} = 0$.

Suppose now $\un{x}$ is a vector in the null-space of $U$, so that  $U\un{x}
=0$. Let $T$ be the $RG$-matrix   with first row $\un{x}$. Now $\un{x}$
determines $T$. Then $UT$ is an $RG$-matrix with first row $0$ and
hence $UT = 0$. Therefore $ut = 0$ in the group ring. Hence $t= qv$
for some $q \in RG$. Therefore $T = QV$. Hence the rows of $T$ are
linear combinations of the rows of $V$. In particular $\un{x}$ is a linear
combination of the rows of $V$. Thus the rows of $V$ generate the
null-space of $U$. Since by linear algebra the null-space of $U$ has
dimension $n-r$ it follows that $V$ has rank $n-r$.
\end{proof}

Suppose $u$ is a zero-divisor in the group ring $RG$, with $G$ 
generated by $g$, and that as {\em polynomials} $uv = g^n -1$,\footnote{Cyclic
codes are defined using such elements.} then $v$ as an element of
$RG$ has the property that it is of minimal degree such that $uv=0$. It
follows that $\rank U + \rank V = n$.
\begin{corollary} If  $uv = g^n -1$ then  $\rank U + \rank V = n$. 
\end{corollary}

\subsection{Rank from cyclic and dihedral zero-divisors}
\slabel{rank-proofs}

Suppose $u$ is a zero-divisor in the group ring $RG$. In the cyclic
group ring case when $\rank u =r$ and $U$ is obtained from the natural
listing of $G$, the first $r$ elements
$u,gu, \ldots , g^{r-1}u$ in the group ring and the 
the first $r$ rows of $U$ give the (full rank) zero-divisor codes.
This is expressed in the following theorem.

\begin{theorem}
\thmlabel{cyclic-independence}
Let $RG$ be a group ring where $R$ is a field and $G$ the cyclic group $\{1,g,g^2,\ldots, g^n\}$. Suppose $u \in RG$ is a zero divisor. Let $r$ be the first value such that $\{u,gu,g^2u,\ldots,g^ru\}$ is linearly dependent.
Then $\rank(u) = r$.
\end{theorem}
\begin{proof}
Linear dependence means that, for some $\alpha_i \in R$, $i=1,\ldots,r-1$ that,
$$
g^r u = \alpha_0 u + \alpha_1 gu + \alpha_2 g^2u + \cdots+ \al_{r-1} g^{r-1} u.
$$
By multiplication by $g^l$,
$$
g^{r+l} u = \alpha_0 g^l u + \alpha_1 g^{1+l}u + \alpha_2 g^{2+l}u + \cdots + \al_{r-1} g^{r-1+l} u.
$$
Therefore for every $x=\di\sum_{i=0}^{n-1} \beta_i g^i\in RG$, $xu$
can be written $\di\sum_{i=0}^{r-1} \beta_i g^i u$, i.e. in terms of $\{u,gu,g^2u,\ldots,g^{r-1}u\}$.
\end{proof}

A similar result holds for the dihedral group.
\begin{theorem}
\thmlabel{dihedral-independence}
Consider the group ring $R D_{2n}$ where $R$ is a field and $D_{2n}$ the
dihedral group. Let $u\in RG$ and $S'$ be the first elements $\{1, b, b^2,
\ldots , b^{k-1} \} \subseteq \{1,b,\dots,b^{n-1}\}$ such that $(S' \union b_k
)u$ is linearly dependent. Subsequently, let $S = S' \union
\{a,ab,\ldots,ab^l\}$ be the first set such that $Su$ is linearly dependent.
Then $\rank(u) = |S|$.
\end{theorem}
\begin{proof}
By the linear dependence,
$$
b^k u = \left(\al_0 + \al_1 b + \cdots + \al_{k-1} b^{k-1}\right) u.
$$
By multiplication on the left on both sides, it can be seen that for every
$k\leq m < n$, $b^m u$ can be written in terms of the previous $k$ powers of $b$
(times $u$) and thus, ultimately, in terms of $S'$. 
Also,
$$
ab^l u = \left(\al_0 + \al_1 b + \cdots + \al_{k-1} b^{k-1} + \beta_0 a +
\beta_1 ab \cdots + \beta_{l-1} ab^{l-1} \right) u.
$$
Multiplication on the left by $b^{l-m}$ yields that 
\begin{align*}
ab^m u = (\al_0 b^{l-m} &+ \al_1 b^{l-m+1} + \cdots + \al_{k-1} b^{l-m+k-1}
\\ 
&+ \beta_0 ab^{m-l}  + \beta_1 ab^{m-l+1} +\cdots + \beta_{l-1} ab^{m-1}) u.
\end{align*}
Thus, for each $l \leq m < n$, $ab^m u$ can be written in terms of the form $b^i
u$ (which we showed above can be written in terms of the first $k$ powers of
$b$), and in terms of the previous $l$ elements of the form $ab^i$.
\end{proof}

\section{Conclusions}
\slabel{conclusions}
We have described a method for producing codes from units and zero-divisors in
group rings, including, vitally, a method to obtain generator and check
matrices for them.

It is our belief that this is an intuitive method which has already shown
promise even in the example codes presented. The framework provides a basis to
anyone wishing to further develop or investigate these codes. The rich synergy
of matrix algebra and group ring structure shows itself to be a powerful new
method.

We note that it is possible to obtain
convolutional codes from unit-derived and zero-divisor codes from group rings.
When a ring $R$ has zero-divisors then it is possible to obtain units in a group
ring $RG$ when the group $G$ is infinite, such as $G=C_\infty$ (the infinite
cyclic group), and from these units to derive convolutional-type codes. Results
of this will appear as future work.

The underlying algebraic structure of group rings often
allows the calculation of distance directly. The description as an
encoding enables one at times to work within the group ring to, for example,
calculate the minimum distance of a code or to make a code with a certain
minimum distance.

Also of note is that codes over the integers can be constructed using known
units in $\Z G$. These include {\em alternating units, Bass units,
Hoeschmann Units, bicyclic units} and others, the details of which may be found
in \cite{milies_grouprings}. Other future work we envisage includes the
generalisation to consider combinations of group ring encodings acting in unison
to produce a code.

\bibliographystyle{siam}
\bibliography{IEEEabrv,codes}

\begin{thebibliography}{10}

\bibitem{bazzi03-groupactions}
{\sc Louay~M.J. Bazzi and Sanjoy~K. Mitter}, {\em Some constructions of codes
  from group actions}, Preprint. Under Submission,  (2003).

\bibitem{blahut03algebaic-codes}
{\sc R.~E. Blahut}, {\em Algebraic codes for data transmission}, Cambridge
  University Press, 2003.

\bibitem{davis79-circulantmatrices}
{\sc P.~J. Davis}, {\em Circulant Matrices}, John Wiley and Sons, Inc., New
  York, 1979.

\bibitem{assmus_reedmuller}
{\sc Jr~E.~F.~Assmus}, {\em On {B}erman's characterization of the
  {R}eed-{M}uller codes}, Journal of Statistical Planning and Inference, 56
  (1996), pp.~17--21.

\bibitem{huffmann03fundamentals_ecc}
{\sc W.~C. Huffman and V.~Pless}, {\em Fundamentals of error-correcting codes},
  Cambridge Univ. Press, 2003.

\bibitem{hughes_2000}
{\sc G.~Hughes}, {\em Constacyclic codes, cocycles and a u+v|u-v construction},
  {IEEE} Trans. Inform. Theory, 46 (2000), pp.~674--680.

\bibitem{hurley_grouprings2006}
{\sc Ted Hurley}, {\em Group rings and rings of matrices}, Inter. J. Pure \&
  Appl. Math.,  (2006), pp.~319--335.

\bibitem{macwilliams69-codesfromideals}
{\sc F.~J. MacWilliams}, {\em Codes and ideals in group algebras},
  Combinatorial Mathematics and its Applications,  (1969), pp.~312--328.

\bibitem{macwilliams71orthogonalcirculant}
\leavevmode\vrule height 2pt depth -1.6pt width 23pt, {\em Orthogonal circulant
  matrices over finite fields, and how to find them}, J. Comb. Theory, 10
  (1971), pp.~1--17.

\bibitem{macwilliams_sloane}
{\sc F.~J. MacWilliams and N.J.A. Sloane}, {\em The Theory of Error-Correcting
  codes}, North-Holland Amsterdam/London/New York/Tokyo, 1998.

\bibitem{milenkovic04blockcirculantldpc}
{\sc O.~Milenkovic, I.~Djordjevic, and B.~Vasic}, {\em Block-circulant
  low-density parity-check codes for optical communication systems}, IEEE
  Journal of Selected Topics in Quantum Electronics, 10 (2004), pp.~294--299.

\bibitem{milies_grouprings}
{\sc C\'esar~P. Milies and Sudarshan~K Sehgal}, {\em An Introduction to Group
  Rings}, Klumer, Dordrecht/Boston/London, 2002.

\bibitem{tang04-algebraicldpc}
{\sc H.~Tan, Jun Xu, Yu~Kou, Shu Lin, and Khaled A.~S. Abdel-Ghaffar}, {\em On
  algebraic construction of gallager and circulant low-density parity-check
  codes}, {IEEE} Trans. Inform. Theory,  (2004), pp.~1269--1279.

\bibitem{tanner2004-qcldpc}
{\sc R.M. Tanner, D.~Sridhara, A.~Sridharan, T.E. Fuja, and Jr.
  Daniel~Costello}, {\em Ldpc block and convolutional codes based on circulant
  matrices}, {IEEE} Trans. Inform. Theory,  (2004), pp.~2966--2984.

\bibitem{ward-qrcdivisibility}
{\sc Harold~N. Ward}, {\em Quadratic residue codes and divisibility, Handbook
  of Coding Theory}, Elsevier, Amsterdam, 1998, ch.~9, pp.~827--870.

\end{thebibliography}

\end{document}